\documentclass[bibyear]{aa}  

\usepackage{graphicx}
\usepackage{txfonts}
\begin{document}

   \title{Possibility of chromospheric back-radiation influencing
          the lithium line formation in Spite plateau stars}

   \subtitle{}

   \author{Y. Takeda\inst{1,2}}

   \institute{$^{1}$National Astronomical Observatory of Japan,
              2-21-1 Osawa, Mitaka, Tokyo 181-8588, Japan\\
              $^{2}$SOKENDAI, The Graduate University for Advanced Studies, 
              2-21-1 Osawa, Mitaka, Tokyo 181-8588, Japan\\
              \email{takeda.yoichi@nao.ac.jp}
              }
   \date{Received December 14, 2018; accepted December 29, 2018}

 
  \abstract
   {
Spectroscopically determined lithium abundances of metal-poor turn-off 
dwarfs are known to be nearly constant (Spite plateau), 
but manifestly lower than the primordial value 
expected from the standard cosmological model. 
However, abundance determination by using conventional model atmospheres 
may not necessarily be correct since the existence of high-temperature 
chromosphere even in very old stars has been confirmed. 
}
   {
The aim of this study is to examine how the extra UV flux possibly irradiated 
from the chromosphere could affect the formation of the Li~{\sc i} 6708 
line, and whether or not its influence might lead to a solution of the Li 
abundance discrepancy.  
   }
   {
A simple model chromosphere of a uniform thin gray slab emitting only 
thermal radiation is assumed, characterized by optical 
thickness and temperature. By taking into account this 
incident radiation in the surface boundary condition, non-LTE calculations 
for neutral Li atoms are carried out in order to see how the equivalent 
widths and the resulting abundances are affected by these parameters.   
   }
   {
If the parameters are appropriately chosen, the 
strength of the Li~{\sc i} 6708 line can be  reduced by a factor 
of $\sim$~2--3 due to overionization caused by enhanced UV radiation, 
leading to an apparent lowering of the abundance by $\sim$~0.3--0.5~dex, 
which is consistent with the discrepancy in question. Moreover, the 
observed slight metallicity-dependent trend of the plateau can also 
be reproduced as a result of the change in atmospheric transparency.
   }
   {
Superficial underestimation of Li abundances due to considerable 
overionization caused by chromospheric radiation may be regarded as 
a ponderable interpretation for the cosmological Li problem. 
The touchstone to verify this model would be to check the existence of 
significantly enhanced UV radiation in these Spite plateau stars,
which should be detected if this scenario is valid, 
although very few such UV spectrophotometric observations have  been 
done to date. 
   }

   \keywords{
             Line: formation --
             Radiative transfer --
             Stars: abundances --
             Stars: atmospheres --
             Stars: chromospheres --
             Stars: Population~II
               }

   \maketitle
%

\section{Introduction}

Spite \& Spite (1982) reported in their pioneering work that the 
surface Li abundances of metal-poor late F and early G dwarfs 
in the halo were almost constant at $A$(Li)~$\sim 2.1$ (logarithmic number
abundances in the usual normalization of $A_{\rm H} = 12.00$) irrespective of 
metallicity.
Although this value was considered at first to represent the initial 
Li composition created by Big-Bang nucleosynthesis (BBN), due to 
its remarkable constancy, it was later revealed to be considerably lower
(by $\sim$~0.4--0.5~dex)  than the cosmologically predicted primordial 
value, when the key parameter (baryon-to-photon ratio) for light-element
production was established by analyzing the power spectrum of 
Cosmic Microwave Background (CMB) based on the balloon-borne {\it BOOMERANG}
experiment (de Bernardis et al. 2002) or the {\it WMAP} observation from space
(Spergel et al. 2007).
This so-called ``cosmological Li problem'' has attracted the attention 
of many astrophysicists, and a number of studies have been published on the subject. 
The historical aspect and the current status of this problem is summarized
in the comprehensive review by Fields (2011), which  includes most of the important
references  of that time. 

Observationally, it has been almost established that the Li abundances 
spectroscopically determined for comparatively hot halo dwarfs near the turn-off point 
(corresponding to an effective temperature of 6500~K$\ga T_{\rm eff} \ga 6000$~K)
are remarkably similar at $\sim$~2.1--2.2 (or  higher by $\sim$~0.1--0.2~dex 
depending on the adopted $T_{\rm eff}$ scale) with a considerably small dispersion 
over the metallicity range of $-3 \la$~[Fe/H]~$\la -1.5$, often called the Spite plateau, 
though this plateau was revealed to show a slight [Fe/H]-dependent trend (decreasing 
tendency with a lowering of metallicity) that begins to breaks down (i.e., 
depletion with rather large dispersion) at the extremely metal-poor regime of 
[Fe/H] below $\sim -3$ (e.g., Mel\'{e}ndez et al. 2010; Sbordone et al. 2010; 
and  references therein). 

On the cosmological side, most people nowadays seem to believe that the primordial 
Li abundance due to BBN has been reliably settled at $A$(Li) = $2.64\pm 0.03$\footnote{
The value of 2.64 is used as the primordial lithium abundance throughout this paper, 
although slightly higher values of $\sim 2.7$ reported by successive studies may 
be more adequate (cf. Coc et al. 2012, 2014).} (Spergel et al. 2007), which is 
the value derived from the CMB observation by {\it WMAP}. 
Accordingly, based on the notion that the Spite plateau  value is not primordial, 
various theoretical interpretations have been proposed to date to account 
for the reason why the observed lithium abundances of very old turn-off 
dwarf stars underwent appreciable changes from the original BBN value, for example  
solutions related to nuclear or particle physics, magnetic separation 
in the early universe, processes related to stellar physics such as depletion 
in the pre-main sequence phase, or gravitational settling coupled with turbulent 
mixing (see  references  in Sect.~3 of Fields 2011 or in Sect.~1 
of Fu  et al. 2015). 
An in-depth discussion given by Asplund et al. (2006) is also informative. 
 
Even so, it may be worthwhile to turn to the more fundamental question, 
``Are the observed Li abundances of Spite plateau stars truly reliable?''
Actually, Kurucz (1995) once raised a critical remark that Li abundances
derived by the conventional method could be considerably underestimated
(by a factor of $\sim 10$) because classical 1D model atmospheres do not
correctly describe the actual 3D inhomogeneous convective structure consisting 
of hot and cool convective cells and the Li~{\sc i} 6708 line is highly
temperature-sensitive under the condition of almost all Li atoms being ionized. 
However, successive theoretical studies by different groups on 
the Li line formation in hydrodynamically inhomogeneous atmosphere 
showed that Kurucz's (1995) claim was not correct because it was 
based on a too simplified, physically unrealistic modeling (Asplund 
et al. 1999; Cayrel \& Steffen 2000; Asplund et al. 2003).
Accordingly, it is now generally believed that neither the 3D effect nor 
the non-LTE effect can solve the Li discrepancy problem. 
This lead Asplund et al. (2006)(Sect. 7.2) to state that
``{\it although the last word has not been said on the representation
of the physics of the stellar atmospheres and the formation
of the lithium lines, it would be extremely surprising if systematic
errors in these areas were shown to resolve the lithium problem.}''

However, our understanding of stellar atmospheres is far from complete;
in particular, little is known about the upper atmospheric layers of
metal-poor stars. In this respect, it should be noted that Takeda \& 
Takada-Hidai (2011a) reported the detection of the He~{\sc i} 10830 line 
in absorption  in disk population~I stars and  in almost all 
population~II stars in a wide range of metallicity down to [Fe/H]~$\sim -3.7$ as well,
which suggests the existence of a high-temperature chromosphere\footnote{
In this paper, the term chromosphere is used broadly to indicate the 
high-temperature zone in the upper atmosphere 
(i.e., not  the layer specifically similar to the solar 
chromosphere). As such, this word may also be regarded to represent
the whole concept including chromosphere, transition layer, and corona 
in the solar analogy.} because such a high-excitation line 
($\chi_{\rm exc} = 19.82$~eV) cannot be observed in late-type stellar 
atmospheres with ever decreasing temperature towards the surface. 
Here, the behavior of line strengths is different for low-gravity giants
(considerably large star-to-star dispersion) and high-gravity dwarfs (almost 
constant strength irrespective of metallicity over $-4 \la$~[Fe/H]~$\la -1$) 
(cf. Fig.~4 therein). While the atmospheric heating for the former giants group 
may be due to some physical mechanism specific to unstable atmospheres (e.g., 
pulsation-induced shock) reflecting the low-density condition, the characteristic 
of the latter dwarfs group was rather unexpected because a rotation-induced dynamo 
mechanism would not work effectively in old stars where rotation should 
have been decelerated. So, this may suggest that rotation-independent 
``basal'' chromospheric activity ubiquitously exists in all late-type dwarfs.   

Whatever the physical process (acoustic wave, local dynamo, etc.) 
responsible for such a basal chromosphere may be, the important point is that  
very metal-poor halo dwarfs should have high-temperature layers ($T \sim 10^{4}$--$10^{5}$~K) 
 in their upper atmospheres. 
If so, it is interesting to check whether the radiation emitted back from 
the hot chromosphere can have a significant influence on the formation 
of the Li~{\sc i}~6708 line (e.g., apparent weakening due to overionization) 
because the fraction of neutral lithium is so small ($<0.1$\% in the photosphere 
of solar-type dwarfs) as to be very vulnerable to a slight change in the 
ionization balance. Could this idea shed any light on the Li-gap problem 
of Spite plateau stars?

This study was planned to examine this possibility by assuming a simple 
parameterized model. The points to be clarified are as follows:
\begin{itemize}
\item
Is it ever feasible to reproduce the Li abundance trend of the Spite plateau 
(i.e., systematic reduction by $\sim$~0.3--0.5 dex from the primordial value) 
under the influence of chromospheric back-radiation by an appropriate 
choice of parameters? 
\item
If so, what about the observational signature expected? Is it possible to 
find a way to verify such a model by actual observations?  
\end{itemize}

\section{Line formation under the existence of chromosphere}

\subsection{Modeling of chromospheric radiation}

   \setcounter{figure}{0}
   \begin{figure}[t]
   \centering
   \includegraphics[width=0.3\textwidth, angle=90]{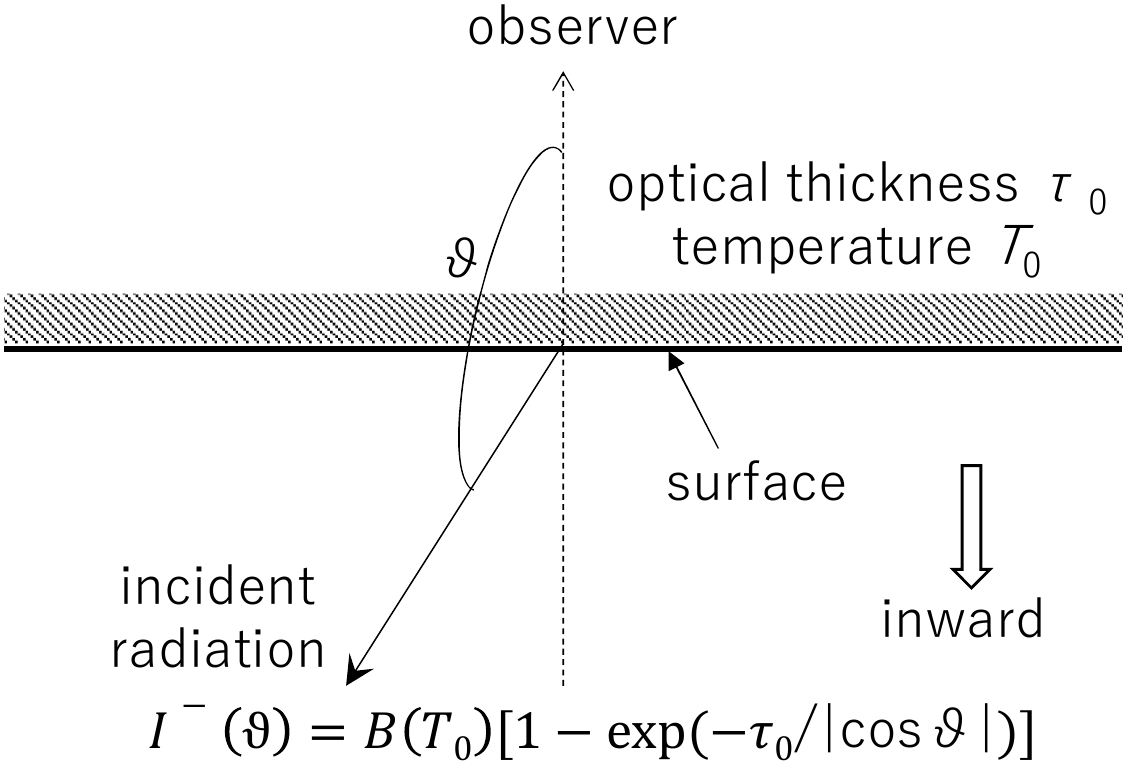}
      \caption{
Schematic description of the simple model chromosphere (uniform gaseous slab
with temperature $T_{0}$ and optical thickness $\tau_{0}$ lying just 
above the atmosphere) emitting thermal radiation incident at the surface.   
}
         \label{schematic}
   \end{figure}

Since very little is known about the nature of chromospheres in old 
metal-poor stars, the following assumptions are made in this study
in order to model the incident chromospheric radiation as simply as possible:\\
-- (1) the chromosphere is modeled by a thin plane-parallel gaseous slab 
lying just above the normal model atmosphere;\\
-- (2) this slab (homogeneous with frequency-independent opacity)
is represented by two parameters (optical thickness $\tau_{0}$ and temperature 
$T_{0}$), and emits only the thermal radiation with the source function  
expressed by the Planck function;\\
-- (3) the structure of underlying the normal model atmosphere is not
affected by the existence of such a high-temperature layer adjacent 
to the surface.

A schematic description of this model is depicted in Fig.~1.
The chromospheric back radiation incident to the surface
is expressed as
\begin{eqnarray}
I^{-}_{\nu}(\theta) = B_{\nu}(T_{0})[1 - \exp(-\tau_{0}/|\cos\theta|)] \nonumber \\
               \simeq  \tau_{0} B_{\nu}(T_{0})/|\cos\theta|  \;\;\; ({\rm for} \; \tau_{0} \ll 1), 
\end{eqnarray} 
where $I^{-}_{\nu}(\theta)$ is the chromospheric radiation (specific intensity) 
incident to the surface at an angle $\theta (> \pi/2)$ relative to the outward normal
and $B_{\nu}$ is the Planck function. 

\subsection{Non-LTE calculation}

Now that the incident chromospheric radiation $I^{-}_{\nu}$ has been specified, 
the next step is to carry out non-LTE calculations (i.e., evaluation of the 
number population for each energy level of neutral lithium as functions of 
depth, by solving the coupled equations of statistical equilibrium and 
radiative transfer) in the same manner as usual, except that $I^{-}_{\nu}(\theta)$ 
given by Eq.~(1) is used as the surface boundary condition instead of 
$I^{-}_{\nu}(\theta) = 0$.

As for the basic model atmospheres, the Kurucz (1993a) ATLAS9 models with 
$T_{\rm eff} = 6250$~K and $\log g = 4.0$ (corresponding to the representative 
atmospheric parameters of Spite plateau stars) but with widely different 
metallicities ([Fe/H] ranging from 0 down to $-5$ with a step of 1~dex) 
were adopted. Similarly, the Kurucz (1993b) opacity distribution function 
corresponding to the model metallicity was used to calculate the 
radiation field at each depth. For the microturbulence and input lithium 
abundance, a typical value of 1.5~km~s$^{-1}$ and the primordial value of 
$A$(Li) = 2.64 (see  footnote~1) were assigned, respectively; 
see Takeda \& Kawanomoto (2005, Sect.~3) and the references therein for 
the details of the adopted non-LTE calculation procedures.\footnote{The 
reasonability of our non-LTE calculation program may be assured by checking 
the consistency of computed non-LTE corrections with those independently 
derived by previous studies using different codes; cf. Takeda \& Kawanomoto 
(2005, Fig.~4) for G dwarfs and Takeda \& Tajitsu (2017, Fig.~8) for G giants.}

Regarding the combination of ($\tau_{0}$, $T_{0}$), which determines the 
incident chromospheric radiation ($I^{-}_{\nu}$), it should be noted that $\tau_{0}$ 
simply scales the strength of $I^{-}_{\nu}$ (as $\propto \tau_{0}$ since $\tau_{0} \ll 1$
should hold), while $T_{0}$ controls both the strength and the energy 
distribution of $I^{-}_{\nu}$. This means that, as far as the UV photoionizing 
radiation field specifically important for Li ionization equilibrium is 
concerned, its strength can be adjusted by changing only $T_{0}$
even if $\tau_{0}$ is fixed at an arbitrary value (tentatively 
assumed to be $10^{-3}$). Given this $\tau_{0}$, four $\log T$ values of 
4.0, 4.3, 4.5, and 5.0 were eventually chosen after some test calculations.
A summary of the adopted parameters for each model is presented in Table~1. 
We note that ``STD'' means the usual non-LTE calculation done on
the standard model without chromospheric irradiation.

\setcounter{table}{0}
\setlength{\tabcolsep}{3pt}
\begin{table*}[h]
\small
\caption{Summary of the adopted parameters for each model.}
\begin{center}
\begin{tabular}{ccccccccl}\hline\hline
Model & $T_{\rm eff}$ & $\log g$ & [Fe/H] & $v_{\rm t}$ & $A^{\rm give}$ & $\log \tau_{0}$ & $\log T_{0}$ & Remark \\  
(1)   & (2)           & (3)      & (4)    & (5)         & (6)     & (7)             & (8)          & (9) \\
\hline
STD  & 6250 & 4.0 & $0, -1, -2, -3, -4, -5$ & 1.5 & 2.64 & $-\infty$ & $\cdots$ & standard model (no chromosphere) \\
tm30T40 & 6250 & 4.0 & $0, -1, -2, -3, -4, -5$ & 1.5 & 2.64 & $-3$ & 4.0 &  \\
tm30T43  & 6250 & 4.0 & $0, -1, -2, -3, -4, -5$ & 1.5 & 2.64 & $-3$ & 4.3 & near  the observed trend \\
tm30T45  & 6250 & 4.0 & $0, -1, -2, -3, -4, -5$ & 1.5 & 2.64 & $-3$ & 4.5 & near  the observed trend \\
tm30T50  & 6250 & 4.0 & $0, -1, -2, -3, -4, -5$ & 1.5 & 2.64 & $-3$ & 5.0 &  \\
\hline
\end{tabular}
\end{center}
(1) Model code. (2) Effective temperature (in K). (3) Logarithmic surface gravity in cgs units 
(in dex). (4) Metallicity represented by the differential logarithmic abundance of Fe 
relative to the Sun (in dex). (5) Microturbulent velocity (in km~s$^{-1}$). (6) Lithium abundance 
(logarithmic number abundance in the usual normalization of $A_{\rm H} = 12.00$) assumed as given, 
which is the primordial value taken from Spergel et al. (2007). This Li abundance is used 
throughout this paper for non-LTE calculations and for the evaluation of non-LTE equivalent 
widths. (7) Logarithmic optical thickness of the assumed chromospheric layer
just above the atmosphere. (8) Logarithmic temperature (in K) of the assumed chromospheric 
layer just above the atmosphere. (9) Specific remark.
\end{table*}

\section{Discussion}

\subsection{Influence of chromospheric irradiation}

Figure~2 shows the runs of radiation field $J_{\nu}$ at 2000~\AA\ 
(important for the photoionization of Li), the  ratio of non-LTE to LTE line opacity  
($\simeq b_{1}$), and the ratio of line source function to Planck function
($\simeq b_{2}/b_{1}$) as a function of depth, which were computed for each treatment 
of chromospheric radiation (characterized by $T_{0}$) on the models of
different metallicities ([Fe/H] = 0, $-2$, and $-4$).
It is evident from Figs.~2a--c that $J_{\nu}$ is progressively enhanced 
over $B_{\nu}$ with increasing $T_{0}$, as naturally expected. 
Although this tendency of $J_{\nu} > B_{\nu}$ in UV has already been observed 
even in the STD model, inward irradiation of sufficiently strong $I^{-}_{\nu}$ 
at the surface can make $J_{\nu}$ outweigh $B_{\nu}$ by orders of magnitude.
This indicates that the overionization effect, which is caused by 
an imbalance between the photoionization rate (determined by $J_{\nu}$) and the 
recombination rate (controlled by $B_{\nu}$ because recombination is electron 
collision process), becomes more important as the chromospheric incident 
radiation is intensified.
Figures~2d--f actually suggest that the non-LTE line opacity of the
Li~{\sc i} 6708 line (resonance line: transition from the ground level
to the first excited level) is considerably reduced  compared to 
the LTE line opacity ($l_{0}^{\rm NLTE}/l_{0}^{\rm LTE} < 1$)
because the ground level is depopulated as a result of overionization.
This effect, being more significant with an increase in $T_{0}$,
acts in the direction of line weakening, and is  
the most important impact of chromospheric radiation upon the
line strength. On the other hand, the influence of changing $T_{0}$ upon 
the line source function is generally insignificant except for the strongest 
irradiation case (Figs.~2g--i).

The spectral distributions of the emergent flux ($H_{\nu}$) at the surface\footnote{
It should be noted that this $H_{\nu}$ is the flux at the ``surface'' just below the
chromospheric layer (indicated by the thick line in Fig.~1). In order to obtain 
the flux observed by an external observer, the radiation emitted by the thin 
chromosphere (dotted line in Fig.~3) has to be added.} 
corresponding to each different $T_{0}$ (computed for the [Fe/H] = $-2$ model) 
are depicted in Fig.~3, where the distribution of assumed chromospheric radiation 
corresponding to each $T_{0}$, $\tau_{0}B_{\nu}(T_{0}) [= 10^{-3} B_{\nu}(T_{0})]$, 
is also shown for comparison.    
We can see from this figure that the surface flux is manifestly enhanced in 
the ultraviolet (UV) region (below $\la 3000$~\AA) in the chromospheric models 
in comparison with the STD model, while the change in the optical region 
($\ga 4000$~\AA) is insignificantly small (except for the T50 model). This means 
that the key to clarifying the existence and nature of a possible chromospheric 
layer is to observe the UV energy distribution. 

   \setcounter{figure}{1}
   \begin{figure*}[t]
   \centering
   \includegraphics[width=0.7\textwidth]{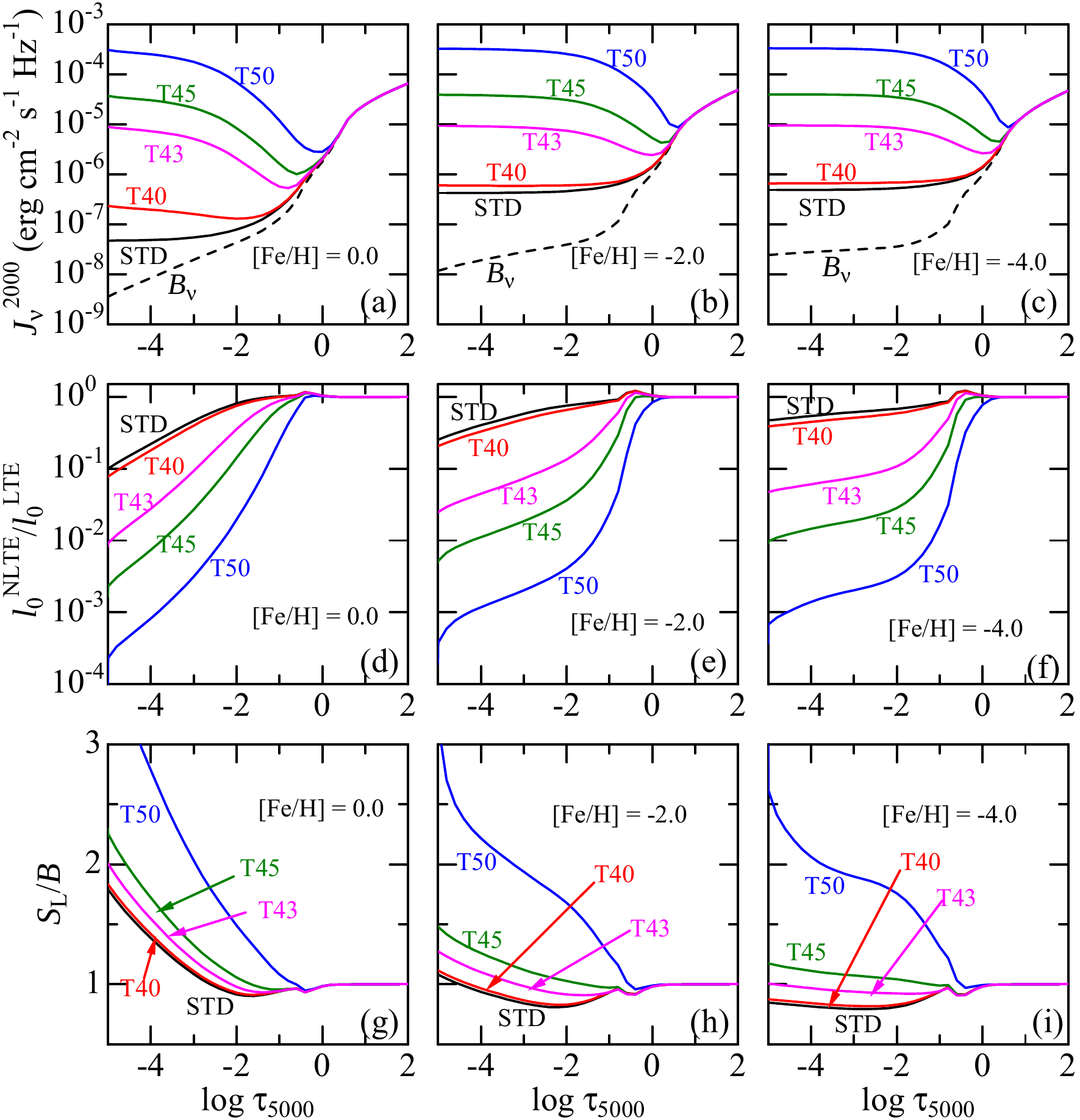}
      \caption{
Top, from left to right  (a, b, c: corresponding to three different metallicities:
[Fe/H] = 0, $-2$, and $-4$): Depth-dependence of 
$J_{\nu}^{2000}$ (mean intensity at $\lambda$ = 2000~\AA) for each of 
the five models (STD, tm30T40, tm30T43, tm30T45, tm30T50);  the local Planck 
function ($B_{\nu}$: dashed line) is also depicted for comparison.
Similarly,  middle  (d, e, f) and  bottom  (g, h, i): 
Runs of $l_{0}^{\rm NLTE}/l_{0}^{\rm LTE}$ (ratio of non-LTE to LTE 
line opacity; $\simeq b_{1}$) and $S_{\rm L}/B$ (ratio of line source 
function to Planck function; $\simeq b_{2}/b_{1}$) of the Li~{\sc i} 6708 line,
respectively. The abscissa for each panel is the logarithm of continuum optical depth 
at 5000~\AA.
              }
         \label{plot_jsb}
   \end{figure*}

   \setcounter{figure}{2}
   \begin{figure}[t]
   \centering
   \includegraphics[width=0.4\textwidth]{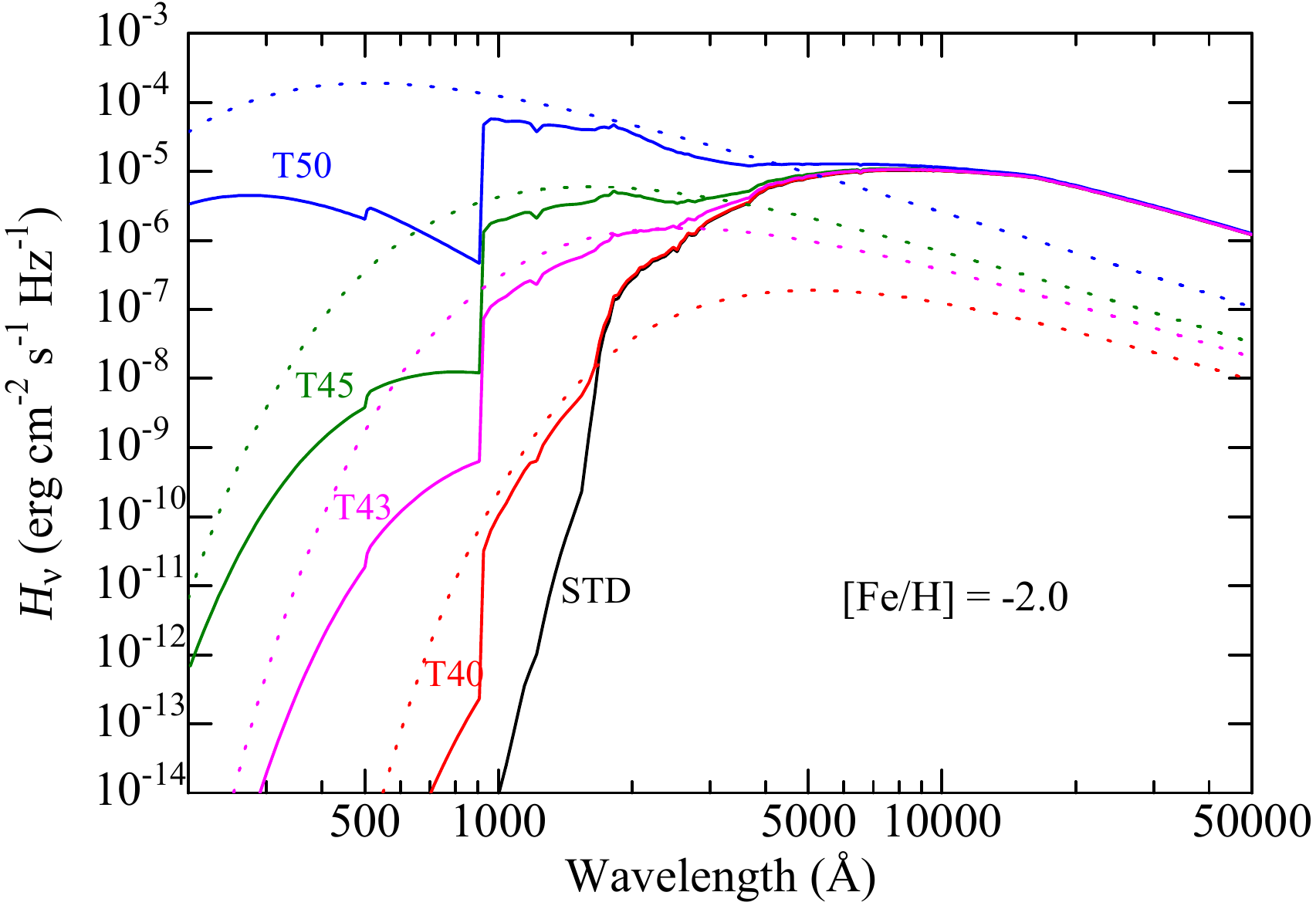}
      \caption{Wavelength distribution of emergent surface flux 
$H_{\nu}$ [$\equiv (1/2)\int_{-1}^{+1} I_{\nu}\mu {\rm d}\mu$, where $\mu = \cos\theta$] 
resulting from each model (solid lines). Wavelength distribution of 
chromospheric radiation [$\tau_{0} B_{\nu}(T_{0}) = 10^{-3}B_{\nu}(T_{0})$]
assumed for each model (dotted lines).
              }
         \label{sedplot}
   \end{figure}

\subsection{Comparison with the observed trends}

We are now ready to discuss the main topic of this study: 
Can the observed trend of the Spite plateau be explained by the line-formation 
models under chromospheric irradiation?

By using the non-LTE departure coefficients resulting from the calculations
in Sect.~2.2, the non-LTE equivalent widths of the Li~{\sc i} 6708 resonance doublet 
line ($W$) were computed for each of the models by using the Kurucz (1993a)
WIDTH9 program,\footnote{This original program was considerably modified
in order to handle the total equivalent width of multicomponent lines
and to take into account the departure from LTE. While 
the non-LTE effect was incorporated through the departure coefficients,
we note that no special change was made to the surface boundary condition of radiative 
transfer equation in this WIDTH9 program (i.e., incident chromospheric radiation 
at the surface was not considered, which is insignificant; cf. footnote~6).}
while assuming the primordial lithium abundance as given ($A^{\rm give} = 2.64$); 
see Table~3 of Takeda \& Tajitsu (2017) for the atomic parameters
of the Li~{\sc i} 6708 line adopted for this evaluation of $W$,
where only lines of $^{7}$Li were included ($^{6}{\rm Li}/^{7}{\rm Li} = 0$). 
Then, the computed $W$\footnote{
The remark given in footnote~4 also applies  to this case.
Calculations of equivalent widths by the WIDTH9 program are based on 
the emergent flux computed at the surface just below the chromosphere.
Therefore, strictly speaking, the derived results (Table~2) 
do not correspond to those observed by an external observer, for which
the radiation from the chromosphere should be added to the surface flux,
resulting in a somewhat weaker equivalent width due to dilution.
However, this effect is insignificantly small for T40, T43, and T45
($\tau_{0} B_{\nu}$ is less than $\la 1/10$ of $H_{\nu}$ 
at the wavelength of Li~{\sc i} 6708 line, though both are comparable 
for T50; see Fig.~3).}
for each model was further analyzed in the usual manner by using the departure 
coefficient of the standard model (STD in Table~1)\footnote{
For the abundance derivation using the STD model, the difference between 
LTE and non-LTE abundances (i.e., non-LTE correction) is so small (a few 
hundredths of a dex at most) that both are almost the same in the practical sense.} 
to inversely derive $A$, which is the apparent lithium abundance
obtained by treating the chromosphere-affected equivalent width
(computed for the given $A^{\rm give}$) by the classical model without chromosphere. 
The resulting values of $W$ and $A$ for each of the models, which are summarized 
in Table~2, are plotted against [Fe/H] in Figs.~4a and 4b, respectively.\footnote{
In this paper, we focus only on the Li~{\sc i} 6708 line 
since almost all investigations have invoked this resonance line 
for Li abundance determinations.
Even so, the weak subordinate line at $\sim 6103.7$~\AA\ 
($\chi_{\rm exc} = 1.85$~eV) can also be employed to derive the 
Li abundances of Spite plateau stars if the data quality is 
sufficiently high (e.g., Asplund et al. 2006).
According to the calculations similarly done for this 
Li~{\sc i} 6104 line, it turned out that the abundance changes 
due to the overionization caused by chromospheric radiation 
are  similar for both of the 6708 and 6104 lines, though the 
effect of abundance reduction is slightly weaker for the latter case; 
e.g., for the T43 model, ($A_{6708}$, $A_{6104}$) are
(2.424, 2.476), (2.363, 2.423), and (2.337, 2.409) 
for [Fe/H] = $-1$, $-2$, and $-3$, respectively.
}

Figure~4a reveals that the strength of Li~{\sc i} 6708 can be considerably 
decreased. The extent of reduction naturally grows as the chromospheric radiation
is enhanced, but the trend is not so much monotonic as  exponential;
i.e., while little change is observed in the T40 model, a  
decrease of a factor of $\sim$~2--3 is found in the T43 or T45 models, 
the T50 model even yielding a reduction as much as $\sim 1/10$.
More interesting is the metallicity dependence. In the STD model,
$W$ tends to show a marginal increase with a decrease in [Fe/H] (especially
from [Fe/H] = 0 to $-1$), which is due to a lowering of continuum
opacity. However, a progressive decrease in $W$ is observed with 
decreasing metallicity  in the models including chromospheric radiation
(except for T40) because the overionization caused by chromospheric 
irradiation penetrates deeper according to the increase of atmospheric 
transparency (i.e., decrease of opacity/metallicity). This situation is illustrated
in Fig.~5, which shows the metallicity dependence of 
$l_{0}^{\rm NLTE}/l_{0}^{\rm LTE}$ and $S_{\rm L}/B$ (though the former 
influence is decisively important, while the latter is insignificant) 
in the deeper atmosphere including the line-forming region.

As a result of this behavior of $W$, the apparent Li abundances ($A$) 
inversely derived from $W$ by using the standard model also show interesting 
characteristics in the context of the Li discrepancy problem in question 
(cf. Fig.~4b):  $A$ is found to be distinctly below the primordial 
value (2.64) by $\sim$~0.3--0.5~dex for the T43 or T45 models. Moreover, 
the slight [Fe/H]-dependent tendency (i.e., a gradual decrease of $A$ toward 
lower [Fe/H]) seen in the chromosphere irradiation cases is worth noting. 
Our models can reproduce  the extent 
of reduction, and  also the metallicity dependence actually observed in 
Spite plateau stars in the range of $-4 \la$~[Fe/H]~$\la -1.5$,
as illustrated in Fig.~6 where the predicted $A$ versus [Fe/H]
relations for T43 and T45 models are compared with the observed data
(taken from four representative studies). 

Accordingly, the hypothesis of superficial weakening of the Li~{\sc i} 6708 
line (due to substantial overionization of neutral lithium caused by 
extra UV radiation from the chromosphere), which was tentatively proposed 
as a solution to the cosmological Li problem of metal-poor turn-off stars, 
stands as a possibility and is worth further investigation. 

\setcounter{table}{1}
\setlength{\tabcolsep}{3pt}
\begin{table*}[h]
\small
\caption{Computed equivalent widths and Li abundances.}
\begin{center}
\begin{tabular}{rcccccccccc}\hline\hline
[Fe/H] & 
$W_{\rm STD}$ & $A_{\rm STD}$ & $W_{40}$ & $A_{40}$ & $W_{43}$ & $A_{43}$ & 
$W_{45}$ & $A_{45}$ & $W_{50}$ & $A_{50}$ \\
(dex) & (m\AA) & (dex) & (m\AA) & (dex) & (m\AA) & (dex) & (m\AA) & (dex) & (m\AA) & (dex) \\
\hline
 & \multicolumn{2}{c}{[STD]} & \multicolumn{2}{c}{[tm30T40]} & \multicolumn{2}{c}{[tm30T43]} 
 & \multicolumn{2}{c}{[tm30T45]} & \multicolumn{2}{c}{[tm30T50]} \\
0.0 & 42.66 & 2.635 & 41.69 & 2.623 & 33.88 & 2.518 & 26.92 & 2.406 & 13.49 & 2.080 \\
$-1.0$ & 47.86 & 2.645 & 45.71 & 2.621 & 30.90 & 2.424 & 19.95 & 2.214 & 7.59 & 1.771 \\
$-2.0$ & 48.98 & 2.641 & 46.77 & 2.617 & 28.18 & 2.363 & 16.98 & 2.125 & 5.37 & 1.607 \\
$-3.0$ & 48.98 & 2.638 & 46.77 & 2.613 & 26.92 & 2.337 & 15.85 & 2.090 & 4.68 & 1.543 \\
$-4.0$ & 48.98 & 2.636 & 46.77 & 2.611 & 26.30 & 2.324 & 15.49 & 2.078 & 4.47 & 1.521 \\
$-5.0$ & 48.98 & 2.636 & 46.77 & 2.611 & 26.30 & 2.325 & 15.49 & 2.078 & 4.37 & 1.511 \\
\hline
\end{tabular}
\end{center}
Non-LTE equivalent width ($W$) of the Li~{\sc i} 6708 line computed for each model of 
six different metallicities with the given Li abundance of $A^{\rm give} = 2.64$, and 
the corresponding non-LTE Li abundance ($A$) obtained by analyzing the derived 
$W$ with the standard STD model.
\end{table*}

   \setcounter{figure}{3}
   \begin{figure}[t]
   \centering
   \includegraphics[width=0.35\textwidth]{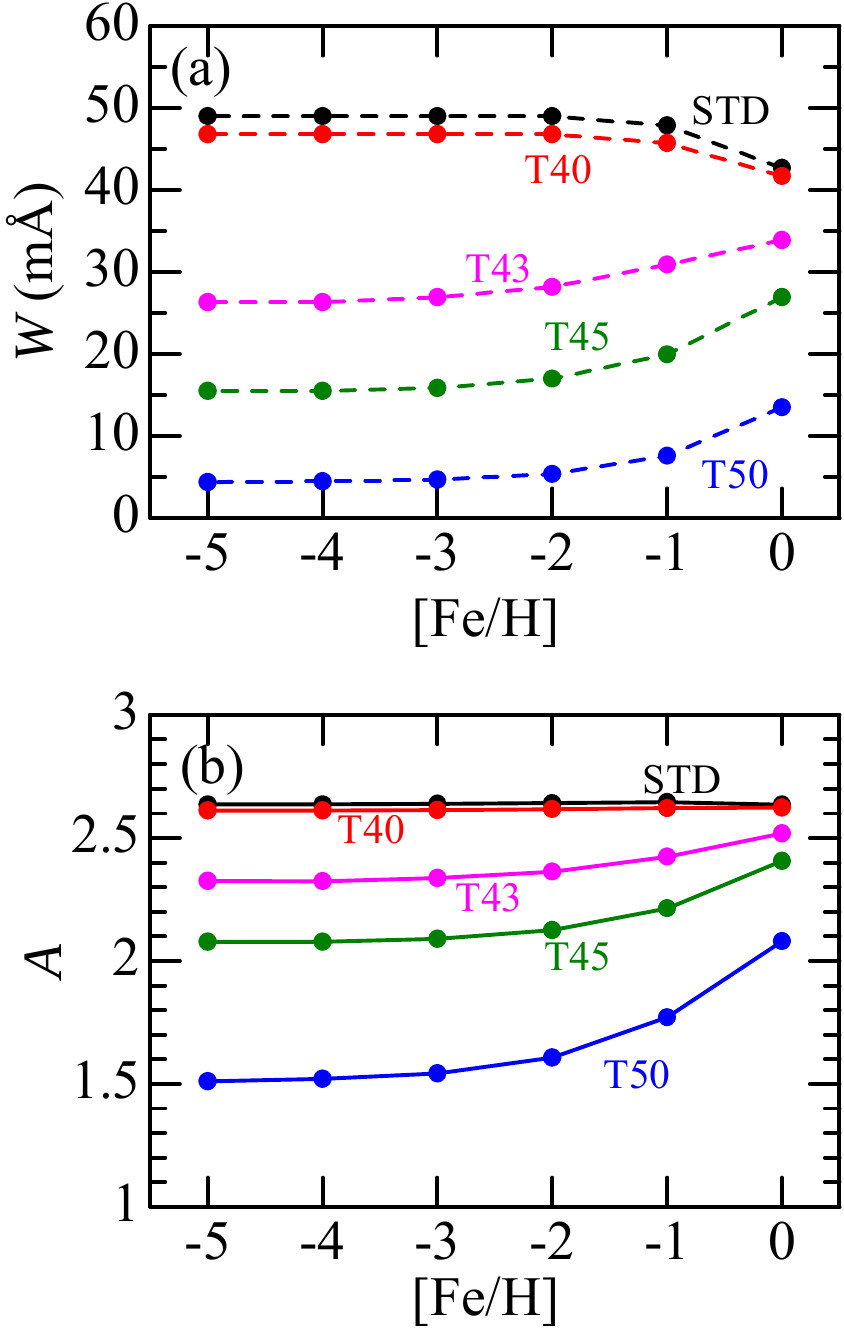}
      \caption{
(a) Non-LTE equivalent widths of Li~{\sc i} 6708 line ($W$; computed for given 
primordial lithium abundance of 2.64 for each of the five models) plotted against [Fe/H].
(b) Non-LTE Li abundances ($A$; inversely obtained from each model's $W$
by using the standard procedure of STD model) plotted against [Fe/H].
              }
         \label{ewabplot}
   \end{figure}

   \setcounter{figure}{4}
   \begin{figure}[t]
   \centering
   \includegraphics[width=0.35\textwidth]{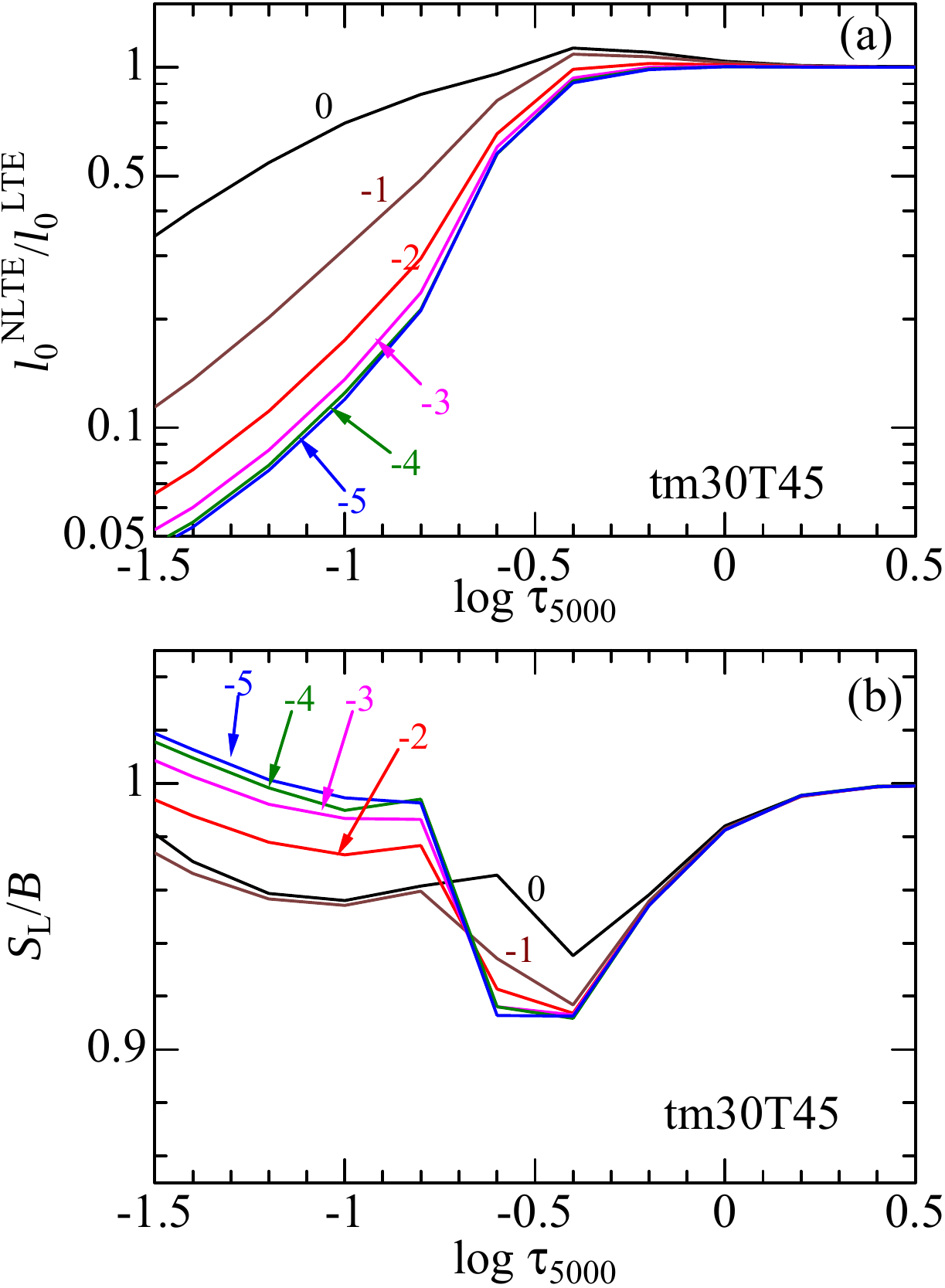}
      \caption{
Depth-dependence of the non-LTE to LTE line opacity ratios (upper panel)
and line source function to Planck function ratios (lower panel),
which were computed for the tm30T45 models of different metallicities
([Fe/H] = 0, $-1$, $-2$, $-3$, $-4$, and $-5$).
              }
         \label{metal_effect}
   \end{figure}

   \setcounter{figure}{5}
   \begin{figure}[t]
   \centering
   \includegraphics[width=0.4\textwidth]{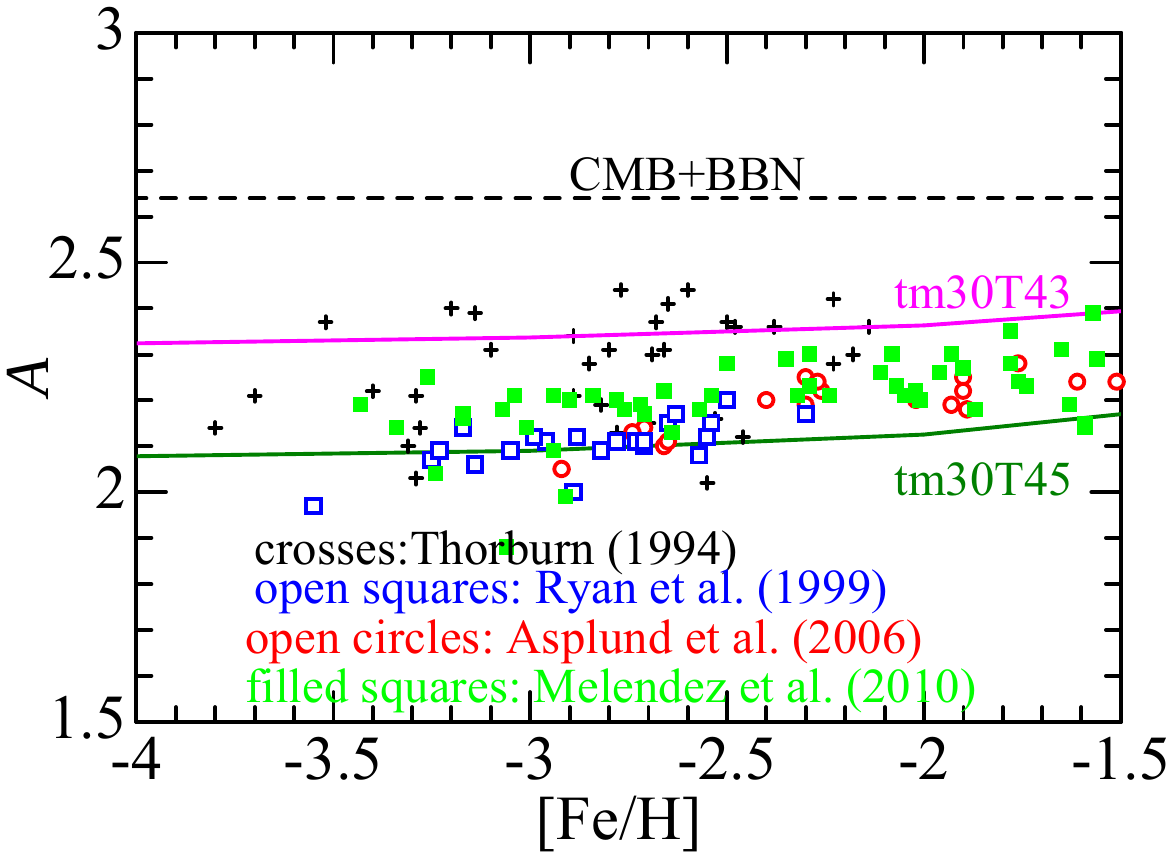}
      \caption{
Comparison of the predicted Li abundance vs. metallicity relations (lines) for 
two models (tm03T43 and tm03T45) with the observed data (symbols) 
of warm metal-poor dwarfs taken from Thorburn (1994) ($T_{\rm eff} > 6100$~K; 
crosses), Ryan et al. (1999) (open squares), Asplund et al. (2006) (open circles),
and Mel\'{e}ndez et al. (2010) ($T_{\rm eff} > 6100$~K; filled squares).
The horizontal dashed line indicates the primordial (CMB+BBN) Li abundance 
of 2.64. 
              }
         \label{compobs}
   \end{figure}

\subsection{The next step}

This pilot study has shown that the characteristics of the Li abundances 
derived for metal-poor turn-off dwarfs, which show considerably small 
dispersion but are systematically lower than the primordial abundance 
established by observational cosmology, could be explained by
the scenario of superficial abundance underestimation as a result of appreciable 
line weakening (due to overionization caused by chromospheric radiation).
However, since the adopted two-parameter model is evidently too simple,
some further considerations (even if speculative) may be due regarding the more 
realistic nature of possible overionization source and the feasibility of
model verification from the observational side.  

Regarding the radiation from the chromosphere, where the region below $\la 2300$~\AA\ 
is the most important in the context of photoionizing neutral lithium atoms, 
the hypothesis of emitting only thermal radiation (continuum spectrum described 
by the Planck function) is probably unrealistic. The spectrum of chromospheric 
radiation must be more complex (including continuum emission as well as 
line emission) as a result of intricate physical processes in the 
high-temperature plasma. However, photoionization by strong UV emission lines 
might be less relevant for the present case involving nearly [Fe/H]-independent 
Li abundances in Spite plateau stars because the strengths of such photoionizing
emission lines would significantly depend upon the metallicity.
Instead, a metallicity-independent mechanism would be  preferable
for producing this radiation. In this context, it may be worth recalling
that the detection of the He~{\sc i} 10830 line corroborated the existence of the chromosphere 
in very metal-poor stars,  which was the motivation for  this study (Sect.~1). 
For the case of the Sun, it is considered that the lower level of this 
He~{\sc i} 10830 transition (2s~$^{3}$S) is mainly populated by  photoionization 
of He~{\sc i} by extreme UV radiation ($\lambda < 504$~\AA) from the solar corona,
 followed by recombination of He~{\sc ii} to populate He~{\sc i}~2s~$^{3}$S
(e.g., Avrett et al. 1994). If  this mechanism is also assumed to be effective 
 for the formation of He~{\sc i} 10830 in population II dwarfs, emission 
of continuum radiation at $\lambda < 2600$~\AA\ is expected as a result of the
He~{\sc ii} + $e$ $\rightarrow$ He~{\sc i}~2s~$^{3}$S recombination process
[$\chi_{\rm ion}$(He~{\sc i}) $-$ $\chi_{\rm exc}$(He~{\sc i}~2s~$^{3}$S) 
= $24.59 - 19.82 = 4.77$~eV (corresponding to $\sim$~2600~\AA], which may contribute
to an efficient photoionization of neutral lithium (whose edge is at $\sim 2300$~\AA).

However, constructing any reliable model of chromsphere (+corona) for simulating
its radiation field is hardly possible for the present case of very metal-poor 
stars since observational information is seriously lacking.
First of all, it is necessary to examine whether a UV photoionizing
radiation strong enough  to cause substantial overionization of Li exists, because spectral 
energy distributions of successful models (which can explain the observed trend 
of Spite plateau) show strong enhancement in the important UV region 
($\sim$~1000--3000~\AA) as remarked in Sect.~3.1 (see Fig.~3).

Unfortunately, such UV photometric observations have barely been done for very 
metal-poor halo dwarfs, for which a small number of available spectroscopic studies 
using the {\it Hubble Space Telescope} paid attention only to the core emission 
of strong lines indicative of chromospheric activity; e.g., Mg~{\sc ii} doublet 
lines at $\sim 2800$~\AA\ or Lyman~$\alpha$ (Peterson \& Schrijver 1997, 2001).
Accordingly, checking the existence of significant UV excess in such comparatively 
warm turn-off dwarfs in the metal-deficient regime ($-4 \la$~[Fe/H]~$\la -1.5$) 
would be a touchstone for deciding the validity of this hypothesis. 
In the case of the Sun, where a wealth of observational data is available, such UV 
excess is not seen. That is, the observed solar flux at $2000 \la \lambda \la 2500$~\AA\
is very low and that at $\lambda \la 2000$~\AA\ is almost zero, and this is consistent 
with the theoretical flux  predicted by the Kurucz (1993a) ATLAS9 solar model atmosphere 
(cf. Takeda et al. 2011b, Fig.~11). Even so, we have no idea how the situation would come out
for the case of very metal-poor stars, which may be markedly different from the Sun
(i.e., metal-rich case where the UV flux is determined by overlapping opacities 
of numerous spectral lines). 

As an alternative possibility, it might be worthwhile to examine the all-sky 
UV photometric survey data of GALEX,\footnote{http://galex.stsci.edu/GR6/} 
which are available in two bands: 
far-UV ($\lambda_{\rm eff} = 1516$~\AA, $\Delta\lambda_{\rm FWHM} = 269$~\AA) and
near-UV ($\lambda_{\rm eff} = 2267$~\AA, $\Delta\lambda_{\rm FWHM} = 616$~\AA).
According to an estimation based on the results in Fig.~3, while $f_{\rm FUV}$ 
is  lower than $f_{\rm NUV}$ by a factor of several tens for the 
case of the STD model, these two fluxes are almost comparable for the T43 
or T45 models. Therefore, if both fluxes were detected at the reliable level 
and successful identification could be made, these data might be usable 
for checking the existence of UV excess.

In any case, it is undoubtedly important to 
observationally establish the UV energy distribution (and also the strength of 
He~{\sc i} 10830 line) for a number of Spite plateau stars,  which  would provide us with 
more insight to the physical nature of chromospheric layers in these stars and 
be able to confirm or disconfirm the interpretation proposed in this study. 
    
Finally, if a special overionization due to extra chromospheric radiation 
were really operative to superficially reduce the Li abundances of
plateau stars by $\sim$~0.4--0.5~dex, its influence should not be limited 
to lithium but should also reach other elements.
For example, the Fe~{\sc i}/Fe~{\sc ii}
ionization equilibrium would be so affected that the abundances derived 
from Fe~{\sc i} lines in LTE are appreciably underestimated, whereas those from 
Fe~{\sc ii} lines remain practically unaffected. While such an overionization 
is already expected in very metal-poor dwarfs even within the framework of
traditional atmospheric models (by $\sim$~0.2--0.3~dex; e.g., Mashonkina et al.
2011), chromospheric irradiation would further exaggerate this effect. 
Although detailed NLTE calculations are necessary for quantitative estimation 
of this impact case by case, it may be expected from a simple analogy that the
resonance lines of other neutral alkali elements with low ionization potential 
(such as Na~{\sc i} 5890/5896 or K~{\sc i} 7665/7699) would behave  
similarly to Li~{\sc i} 6708. Accordingly, it would be interesting
to check whether any meaningful difference exists between the K abundances 
(unaffected by evolution-induced mixing) derived from very metal-poor 
Spite plateau stars and those from cooler red giants at the same metallicity.

\section{Conclusion}

Regarding what is known as the  cosmological Li problem, which is the discrepancy
between the lithium abundances of metal-poor turn-off dwarfs being nearly
constant irrespective of metallicity (Spite plateau) and the primordial
BBN value almost established from the CMB observations (e.g., {\it WMAP}),
various explanations have been proposed, most of which suppose 
that the observed stellar Li abundance reflects the real composition
in the atmosphere and would have been changed (i.e., decreased) from 
the initial value by some physical mechanism. 

This study  casts doubt on this general belief, and proposes that the  
problem might be on the technical side of abundance determination; i.e., 
the surface  Li abundances of these stars might have been underestimated.
This suspicion was motivated by the observational fact that hot chromosphere 
exists in metal-poor dwarfs, as evidenced by the detection of the He~{\sc i} 10830 
line whose strength is almost constant irrespective of the metallicity.
If so, chromospheric UV radiation might induce significant overionization of
neutral lithium and considerable weakening of the Li~{\sc i} 6708 line,
which could lead to an underestimation of the Li abundance if derived by 
the conventional method of analysis. The aim of this investigation was 
to examine this possibility. 

As to the modeling of chromospheric radiation, thermal radiation emitted by 
a uniform slab (characterized by optical thickness $\tau_{0}$ and temperature 
$T_{0}$) was simply assumed. Incorporating this incident radiation in the 
surface boundary condition, non-LTE calculations for neutral 
Li atoms were carried out with different combinations of ($\tau_{0}$, $T_{0}$).
In addition, based on the resulting non-LTE departure coefficients,  
how the equivalent widths and the corresponding abundances are 
affected by these parameters were also investigated.   

The results turned out rather satisfactory. If the parameters are adequately chosen, 
the equivalent width of Li~{\sc i} 6708 can be considerably reduced 
by a factor of $\sim$~2--3 due to the overionization effect caused by an 
enhanced UV radiation irradiated from the chromosphere, which eventually 
leads to an appreciable decrease in the apparent abundance by $\sim$~0.3--0.5~dex,
being consistent with the discrepancy in question. Moreover, the observed slight 
metallicity-dependent slope of the plateau (i.e., Li abundance tends to slightly 
decrease with a decrease in [Fe/H]) can also be reproduced,  because 
the overionization stemming from chromospheric irradiation penetrates deeper with 
an increase in atmospheric transparency (resulting from decreased metallicity). 

Accordingly, the superficial underestimation of Li abundances, which results from an 
appreciable weakening of Li~{\sc i} 6708 line caused by considerable overionization 
due to external radiation from the chromosphere, may be a possible 
interpretation of the cosmological Li problem and is worth further investigation. 
However, since this calculation is based on a simple parameterized model, 
successful reproduction of the observed trend established by arbitrarily 
changing the parameters does not mean that this concept is justified. 
Therefore, in order to check the validity of this hypothesis, it is important 
to confirm the significant UV excess in these Spite plateau stars, 
which should be detected if  significant overionization is actually present,
though very few such UV spectrophotometric observations have  been done to date
for this class of very metal-poor stars. 
   
%

%


\begin{thebibliography}{}
\bibitem[Asplund(2003)]{asplund03}
  Asplund, M., Carlsson, M., \& Botnen, A. V. 2003, A\&A, 399, L31
\bibitem[Asplund(2006)]{asplund06}
  Asplund, M., Lambert, D. L., Nissen, P. E., Primas, F., \& Smith, V. V. 2006, 
  ApJ, 644, 229
\bibitem[Asplund(1999)]{asplund99}
  Asplund, M., Nordlund, \AA, Trampedach, R., \& Stein, R. F. 1999, A\&A, 346, L17
\bibitem[Avrett(1994)]{avrett94}
  Avrett, E. H., Fontenla, J. M., \& Loeser, R. 1994, in Infrared solar physics, 
  Proc.154th IAU Symp. (eds.) D. M. Rabin, J. T. Jefferies, \& C. Lindsey 
  (Dordrecht: Kluwer), 35
\bibitem[Cayrel(2000)]{cayrel00}
  Cayrel, R., \& Steffen, M. 2000, in The Light Elements and their 
  Evolution, Proc. IAU Symp. 198, eds. L. da Silva, M. Spite, 
  J. R. de Medeiros (Astron. Soc. Pacific: San Francisco), 437
\bibitem[Coc(2012)]{coc12}
  Coc, A., Goriely, S., Xu, Y., Saimpert, M., \& Vangioni, E. 
  2012, ApJ, 744, 158
\bibitem[Coc(2014)]{coc14}
  Coc, A., Uzan, J.-P., \& Vangioni, E. 2014, AIP Conf. Ser. Vol. 1594, 
  Can mirror matter solve the cosmological lithium problem? 
  (New York: Am. Inst. Phys.), 12
\bibitem[deBernardis(2002)]{debernardis02}
  de Bernardis, P., Ade, P. A. R., Bock, J. J., et al. 2002, ApJ, 564, 559
\bibitem[Fields(2011)]{fields11}
  Fields, B. D. 2011, Ann. Rev. Nucl. Part. Sci., 61, 47 [also available at 
\texttt{https://ned.ipac.caltech.edu/\\
level5/Sept15/Fields/Fields\_contents.html}]
\bibitem[Fu(2015)]{fu15}
  Fu, X., Bressan, A., Molaro, P., \& Marigo, P. 2015, MNRAS, 452, 3256
\bibitem[Kurucz(1993a)]{kurucz93a}
  Kurucz, R. L. 1993a, Kurucz CD-ROM, No. 13 
  (Harvard-Smithsonian Center for Astrophysics) 
\bibitem[Kurucz(1993b)]{kurucz93b}
  Kurucz, R. L. 1993b, Kurucz CD-ROM, No. 14 
  (Harvard-Smithsonian Center for Astrophysics) 
\bibitem[Kurucz(1995)]{kurucz95}
  Kurucz, R. L. 1995, ApJ, 452, 102
\bibitem[Mashonkina(2011)]{mashonkina11}
  Mashonkina, L., Gehren, T., Shi, J.-R., Korn, A. J., \& Grupp, F. 
  2011, A\&A, 528, A87
\bibitem[Melendez(2010)]{melendez10}
  Mel\'{e}ndez, J., Casagrande, L., Ram\'{\i}rez, I., Asplund, M., 
  \& Schuster, W. J. 2010, A\&A, 515, L3
\bibitem[Peterson(1997)]{peterson97}
  Peterson, R. C., \& Schrijver, C. J. 1997, ApJ, 480, L47
\bibitem[Peterson(2001)]{peterson01}
  Peterson, R. C., \& Schrijver, C. J. 2001, 
  in The 11th Cool Stars, Stellar Systems and the Sun, 
  ASP Conf. Ser. Vol. 223, eds. R. J. Garc\'{\i}a L\'{o}pez, 
  R. Rebolo, \& M. R. Zapatero (San Francisco: Astronomical
  Society of the Pacific), 300
\bibitem[Ryan(1999)]{ryan99}
  Ryan, S. G., Norris, S. E., \& Beers, T. G. 1999, ApJ, 523, 654
\bibitem[Sbordone(2010)]{sbordone10}
  Sbordone, L., Bonifacio, P., Caffau, E., et al. 2010, A\&A, 522, A26
\bibitem[Spergel(2007)]{spergel07}
  Spergel, D. N., Bean, R., Dor\'{e}, O., et al. 2007, ApJS, 170, 377
\bibitem[Spite(1982)]{spite82}
  Spite, F., \& Spite, M. 1982, A\&A, 115, 357
\bibitem[Takeda(2005)]{takeda05}
  Takeda, Y., \& Kawanomoto, S. 2005, PASJ, 57, 45
\bibitem[Takeda(2017)]{takeda17}
  Takeda, Y., \& Tajitsu, A. 2017, PASJ, 69, 74
\bibitem[Takeda(2011b)]{takeda11b}
  Takeda, Y., Tajitsu, A., Honda, S., et al. 2011b, PASJ, 63, 697
\bibitem[Takeda(2011a)]{takeda11a}
  Takeda, Y., \& Takada-Hidai, M. 2011a, PASJ, 63, 547
\bibitem[Thorburn(1994)]{thorburn94}
  Thorburn, J. A. 1994, ApJ, 421, 318
\end{thebibliography}
\end{document}